# Neural Network-Based Processing and Reconstruction of Compromised Biophotonic Image Data

*Michael John Fanous[a], Paloma Casteleiro Costa[a], Çağatay Işıl[a,b,c], Luzhe Huang[a,b,c] and Aydogan Ozcan[a,b,c,d]*

[a] *Electrical and Computer Engineering Department, University of California, Los Angeles 90095 CA, USA*
[b] *Bioengineering Department, University of California, Los Angeles 90095 CA, USA*
[c] *California NanoSystems Institute (CNSI), University of California, Los Angeles 90095 CA, USA*
[d] *Department of Surgery, David Geffen School of Medicine, University of California, Los Angeles 90095 CA, USA*

**Abstract**. In recent years, the integration of deep learning techniques with biophotonic setups has opened new horizons in bioimaging. A compelling trend in this field involves deliberately compromising certain measurement metrics to engineer better bioimaging tools in terms of e.g., cost, speed, and form-factor, followed by compensating for the resulting defects through the utilization of deep learning models trained on a large amount of ideal, superior or alternative data. This strategic approach has found increasing popularity due to its potential to enhance various aspects of biophotonic imaging. One of the primary motivations for employing this strategy is the pursuit of higher temporal resolution or increased imaging speed, critical for capturing fine dynamic biological processes. Additionally, this approach offers the prospect of simplifying hardware requirements and complexities, thereby making advanced imaging standards more accessible in terms of cost and/or size. This article provides an in-depth review of the diverse measurement aspects that researchers intentionally impair in their biophotonic setups, including the point spread function (PSF), signal-to-noise ratio (SNR), sampling density, and pixel resolution. By deliberately compromising these metrics, researchers aim to not only recuperate them through the application of deep learning networks, but also bolster in return other crucial parameters, such as the field-of-view (FOV), depth-of-field (DOF), and space-bandwidth product (SBP). Throughout this article, we discuss various biophotonic methods that have successfully employed this strategic approach. These techniques span a wide range of applications and showcase the versatility and effectiveness of deep learning in the context of compromised biophotonic data. Finally, by offering our perspectives on the exciting future possibilities of this rapidly evolving concept, we hope to motivate our readers from various disciplines to explore novel ways of balancing hardware compromises with compensation via artificial intelligence (AI).

## 1 Introduction

The integration of deep learning with biophotonic technologies[1,2] heralds an unprecedented era in imaging and microscopy, characterized by transformative enhancements in the realm of image reconstruction[3,4]. Central to this innovative shift is the concept of neural network-based data processing, a powerful approach that has gained significant traction within the field of bioimaging. Neural network compensation hinges on a deliberate strategic compromise – a calculated choice to sacrifice certain measurement metrics in exchange for their later restoration or enhancement through the application of deep learning models trained on a substantial amount of data. This strategic trade-off can serve different purposes, including increasing temporal resolution and imaging speed, simplifying hardware configurations, and reducing costs. It also introduces novel ways to deal with typical peripheral bioimaging issues, such as phototoxicity[5] and photobleaching[6,7], a concern when dealing with sensitive biological specimens.

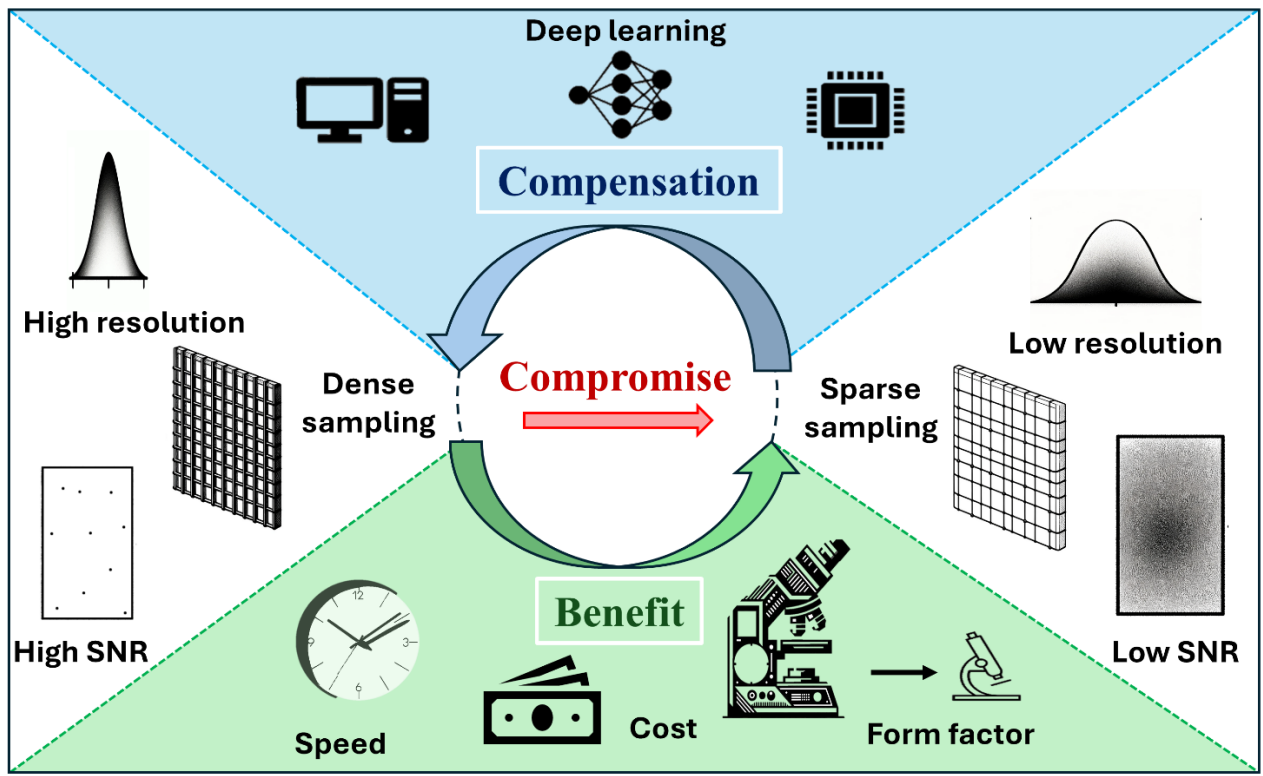

**Figure 1. Schematic illustrating the concept of neural network-based image processing and reconstruction of compromised photonic data in terms of resolution, sampling density, and SNR. Deep learning compensation results in speed, cost, and/or size benefits.**

In this article, we cover a multitude of imaging systems[8-21] that involve deliberate impairments, including to the Point Spread Function (PSF), Signal-to-Noise Ratio (SNR), sampling volume, and pixel resolution, which are recuperated along with enhancements to one or more of the following: spatial/temporal resolution, field-of-view (FOV), depth-of-field (DOF) and space-bandwidth product (SBP). These intricately calculated trade-offs, while necessitating the initial relinquishment of specific metrics, can offer significant practical benefits through the application of deep learning-based inference, as illustrated in **Figure 1**.

Our exploration encompasses a review of over a dozen such biophotonic approaches, each of which skillfully leverages the compensatory capacities of deep learning[8-21]. These endeavors show how artificial intelligence (AI) can help overcome a variety of bioimaging challenges, furthering the field of biophotonics. **Table 1** presents some of the main articles that are covered in this manuscript and highlights the *compromised/compensated* metrics involved in each case. Our extensive review is an attempt to delineate the synergistic relationship that exists between deep learning and biophotonics and is divided into three sections: (i) *refocusing*, (ii) *reconstruction with less data*, and (iii) *improving image quality and throughput.* Though there is some overlap among methods for each section, as indicated by the symbols listed in **Table 1**, we assigned each technique to the category most pertinent to the metrics that are conceded and subsequently restored or enhanced. Each segment comprises a few representative major studies that help illustrate the powerful assistance that deep learning can lend in advancing biophotonic technologies.

**Table 1. Neural network-based image processing and reconstruction methods using compromised data with the corresponding trade-offs and overall benefits**

| Method | Authors | Network architecture | Compromised | | | | Enhanced/Restored | | | |
|---|---|---|---|---|---|---|---|---|---|---|
| | | | PSF/NA | Sampling | SNR | Pixel resolution | Spatial resolution | FOV | DOF | SBP |
| ● Deep-R | (Luo et al. 2021)[8] | GAN[22] | ✗ | | | | | | ✓ | |
| ▲ ● W-NET | (Yang et al. 2021)[9] | Two cascaded U-Nets (with a GAN-based loss) | ✗ | | ✗ | | ✓ | | ✓ | ✓ |
| ● GANscan | (Fanous and Popescu 2022)[10] | GAN | ✗ | | | | ✓ | ✓ | | ✓ |
| ● Single shot autofocus microscopy | (Pinkard et al. 2019)[20] | Fully connected Fourier Neural Network (FCFNN) | ✗ | | | | ✓ | | | |
| ● ◆ eFIN | (Chen et al. 2023)[11] | Enhanced Fourier Imager Network | ✗ | ✗ | | ✗ | ✓ | | ✓ | ✓ |
| ◆ Recurrent-MZ | (Huang et al. 2021)[12] | Convolutional recurrent network[23] | | ✗ | | | | | ✓ | |
| ◆ SS-OCT | (Zhang, Liu et al. 2021)[14] | U-Net | | ✗ | | | ✓ | | | |
| ◆ Few shot holographic reconstruction | (Huang et al. 2022)[13] | Recurrent U-net (with a GAN-based loss) | | ✗ | | ✗ | ✓ | | | |
| ● ◆ Single-shot Fourier ptychography | (Cheng et al. 2019)[15] | CNN with residual blocks | ✗ | | | | ✓ | | | ✓ |
| ▲ ● Mobile phone microscopy | (Rivenson et al. 2018)[16] | CNN with residual blocks | ✗ | | ✗ | | ✓ | ✓ | | ✓ |
| ▲ ◆ Low light SIM | (Jin et al. 2020)[18] | U-Net | | ✗ | ✗ | | ✓ | | | |
| ▲ Fast STED | (Ebrahimi et al. 2023)[21] | UNet-RCAN[24] | | | ✗ | | ✓ | | | |
| ▲ Denoising SRS | (Manifold et al. 2019)[19] | U-Net | | | ✗ | | ✓ | | | |
| ◆ ▲ Overlapped microscopy | (Yao at al. 2022)[17] | CNN | | | ✗ | | | ✓ | | ✓ |

● *Refocusing;* ◆ *Reconstruction with less data;* ▲ *Improving image quality and throughput*

## 2 Refocusing

The process of obtaining high-fidelity, all-in-focus images is crucial in the analysis of biophotonic data. Traditional refocusing methods, as illustrated in **Figure 2a**, often rely on mechanical scanning techniques wherein multiple images are captured at different focal planes in a serial 'stop-and-stare' fashion. These images are then algorithmically analyzed to identify the best in-focus image and the focal position. This process requires extensive data acquisition and processing time. Transitioning from mechanical to computational refocusing (**Figure 2b)** is enabled using neural network-based refocusing approaches.

One of the most important metrics for adequate focusing that has seen successful compromises and improvements through deep learning is the PSF, which characterizes the spatial extent of a point source in the acquired image, and its enhancement is pivotal for achieving sharper, more detailed images[25]. An out-of-focus or blurry image thus corresponds to a situation where the PSF is in some way enlarged and distorted. Deep learning has already demonstrated its ability to enhance the PSF[9,26-28], thus improving high-resolution imaging. In the specific context of neural network compensation, numerous studies have already made significant strides[8-11,26]. These works, harnessing the power of deep learning, have not only managed to accelerate or greatly facilitate the imaging process, but have also expanded the capabilities of microscopy systems to capture finer details and provide crisper, higher quality images. In this section, we discuss various leading methods of PSF engineering and refinement, leveraging AI in fluorescence and brightfield microscopy, holography, and phase contrast microscopy.

One such technique is the single-shot autofocusing method termed Deep-R[8]. In this method, offline autofocusing[29] is rapidly and blindly achieved for single-shot fluorescence and brightfield microscopy images acquired at arbitrary out-of-focus planes (**Figure 2c**). Deep-R significantly accelerates the autofocusing process (about 15-fold faster) using an automated focused image inference, and all without any hardware modifications or the need for prior knowledge on defocus distances. A similar operation is accomplished with the network termed W-Net[9], which comprises a cascaded neural network and a double helix PSF[30], representing a noteworthy advancement in the context of virtual refocusing and consequential enhancement of the DOF. This deep learning-based offline autofocusing approach enhances the quality of image reconstruction, while extending the DOF by approximately 20-fold. The W-Net model was developed as a sequence of two neural networks designed to enhance image quality through computational refocusing and reconstruction. The first part of the model works on adjusting a PSF-engineered input image to target specific planes within the sample volume. Following this initial calibration, the second network takes over, utilizing the virtually refocused images to conduct a comprehensive image reconstruction, as shown in **Figure 2d**. This process is guided by a cross-modality transformation (wide-field to confocal)[31], ultimately producing images that are comparable in quality to those obtained from confocal fluorescence microscopy. One way to view and combine these two approaches is: *a compromised PSF is exploited for purposes of speed and simplicity, and then rectified using AI.*

Deep learning-enabled refocusing has also been extensively demonstrated in holographic microscopy imaging[32-40]. Among such models is the enhanced Fourier Imager Network (eFIN) framework[11]. eFIN is a highly versatile solution for simultaneous hologram reconstruction, pixel super-resolution, and image autofocusing. eFIN enables sharper and higher-resolution imaging while maintaining image quality and is designed for both phase retrieval and holographic image enhancement on low-resolution raw holograms through its inference process. Building on the foundational Fourier Imager Network (FIN)[41]- a network that achieves better hologram reconstruction than convolutional neural networks (CNNs) by synergistically utilizing both the spatial features and the spatial frequency distribution of its inputs - eFIN showcases notable advancements in its model structure, particularly through the incorporation of a simplified U-Net within its Dynamic Spatial Fourier Transform (SPAF) module, which utilizes input-dependent kernels and better adapts to inputs with varying features. This innovative architecture enables eFIN to seamlessly combine pixel super-resolution with autofocusing functionalities within a singular framework. A distinctive feature of eFIN is its proficiency in autofocusing across an extensive axial range of ± 350 μm, coupled with its remarkable ability to accurately estimate the axial positions of input holograms by leveraging physics-informed learning techniques[34], thereby eliminating the reliance on actual axial distance measurements.

An altogether different kind of deep learning enhancement of the PSF is found in the rapid brightfield and phase contrast scanning method known as GANscan[10]. This powerful approach harnesses generative adversarial networks (GANs)[22] to restore sharpness of images extracted from motion-blurred videos (**Figure 2e**). In this case, the PSF is elongated horizontally and narrows the horizontal spatial frequencies. Adjusting for this defect, GANscan enables ultra-fast image acquisition through motion-blurred scanning. The resulting acquisition rate matches the leading-edge Time Delay Integration (TDI)[42] technology's performance, achieving 1.7–1.9 gigapixels within 100 seconds. Such a technique offers an efficient and cost-effective way to accomplish rapid digital pathology scanning using only basic optical microscopy hardware.

Another approach[20] demonstrates the same end goal of autofocusing with a phase contrast modality, but in an indirect manner. Unlike the methods previously mentioned, this technique employs a neural network to predict the focal offset, which then prompts a change to the optical hardware. The PSF manipulation is in this case *mechanical* in nature and only the result of an AI-based inference and successive adjustment. Using just one or a few off-axis LEDs, the method allows for a significant speedup in obtaining in-focus images, a critical factor for accurately capturing dynamic biological processes in real time. The 'fully connected Fourier neural network (FCFNN)' employed in this technique is designed to exploit the sharp features resulting from coherent illumination, allowing it to make accurate focus predictions from a single image (**Figure 1f**). This concept aligns closely with some of the recent advancements in digital holography, where similar principles have been employed for rapid, post-experimental digital refocusing[43].

These exemplary instances collectively underscore the transformative power of deep learning in microscopic image refocusing, a critical factor in the pursuit of high-resolution imaging within the domain of biomedical imaging. By leveraging neural network compensation, these approaches speed up imaging and enhance microscopy systems, resulting in faster acquisition of sharper, high-definition images with better DOF.

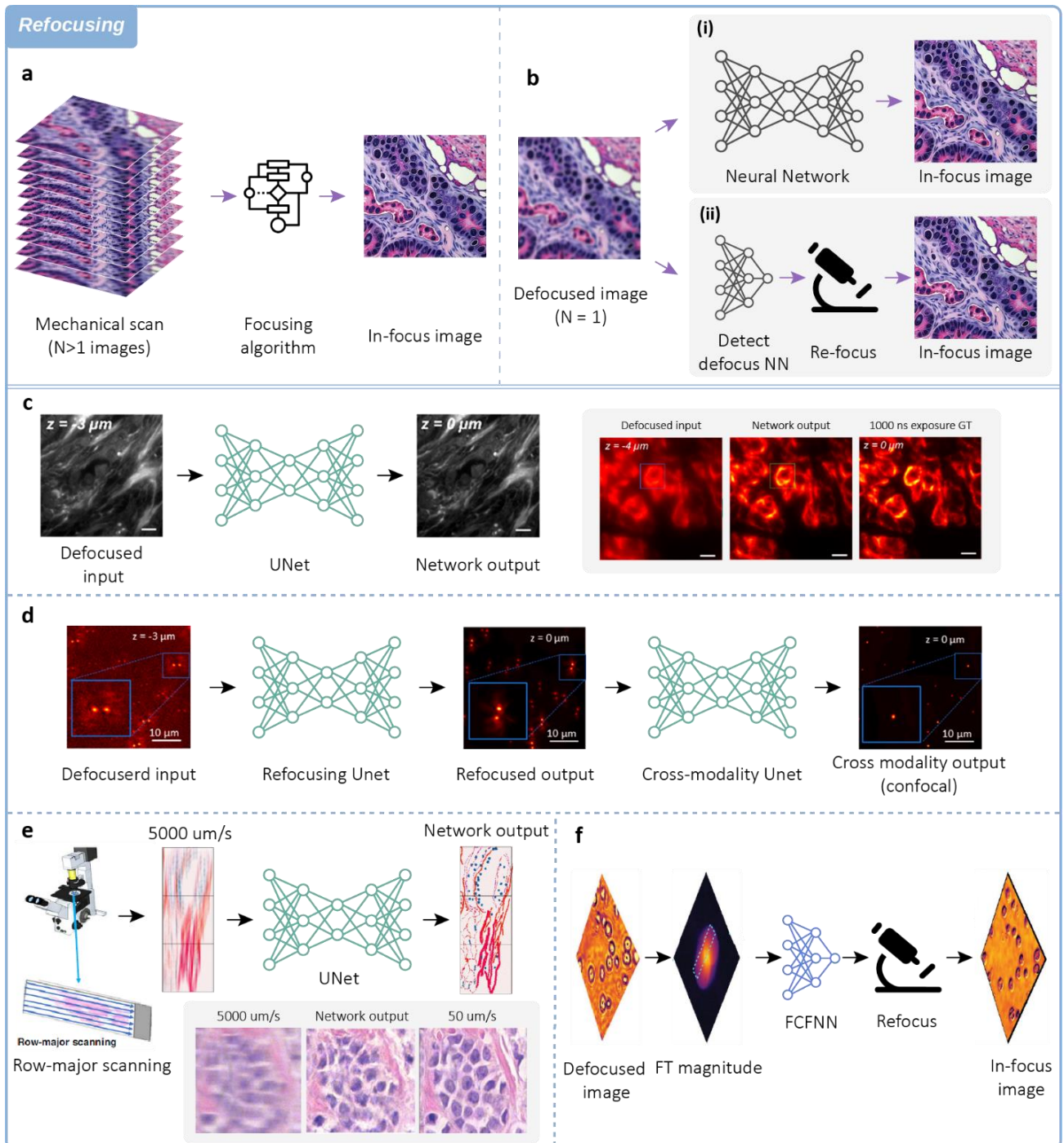

**Figure 2 Refocusing Methods. a** Traditional mechanical scanning technique where multiple defocused images are captured at different focal planes (N>1 images). These images are then analyzed to identify the sharpest in-focus image, requiring extensive data acquisition and processing time**. b** Deep neural network-based approaches to image refocusing (computational or optical). **c** Deep-R blindly autofocuses a defocused image after its capture[8]. **d** Structure of W-Net, containing two cascaded neural networks: (1) virtual image refocusing network and (2) cross-modality image transformation network optimized for DH-PSF[9]. **e** The GANscan method resolves deliberately motion blurred scans that save on time using models trained with relatively slow scans[10]. **f** The FCFNN[20] model uses just one coherent out-of-focus image, which is then analyzed through an established pipeline to acquire a focus prediction, following which the microscope's optics are mechanically adjusted.

## 3 Reconstruction with less data

The quest for efficient image reconstruction in biophotonics often grapples with the challenge of data scarcity. In order to obtain high quality images, especially for three-dimensional (3D) or quantitative systems, a large volume of measurements needs to be acquired, which entails longer imaging durations, more data, and in the case of live biological specimens, exacerbation of problems such as photoxicity[5] and photobleaching[7].

Purposely undersampling measurement data is thus a critical strategy in deep learning-enhanced biophotonics. This concept involves deliberately reducing the amount of data acquired during the imaging process, often entailing certain compromises in measurement metrics. However, the emerging trade-offs allow for various advantages, including increased imaging speed, reduced data acquisition requirements, and minimized photodamage to delicate samples. In this section, we explore how this practice has successfully been applied across a range of modalities, including Fourier ptychography, 3D fluorescence microscopy, optical coherence tomography (OCT), and digital holography, highlighting the diverse applications of this approach in enhancing biophotonic imaging with relatively sparse data.

Typical image reconstruction processes require many input acquisitions that are then fed into an algorithm to generate a decent result (**Figure 3a**). Employing a specially trained deep learning network, however, one may begin with a limited set of input data, a significant reduction from the traditional multi-layered stack, allowing for the minimization of initial data requirements without a meaningful loss in image quality (**Figure 3b**). An example that applies this idea is the single-shot Fourier ptychographic microscopy method[15], which introduces an important approach of strategically undersampling data and employing neural network compensation to nonetheless achieve high-resolution image reconstruction. Fourier ptychography is a computational imaging technique that enables high-resolution, wide-field imaging beyond the single-shot numerical aperture of the optical system employed. This method reconstructs a high-resolution image by stitching together information from a series of low-resolution images taken at different illumination angles[44]. Fourier ptychographic microscopy traditionally requires illuminating and capturing images from multiple LEDs in an array sequentially[45]. However, recent innovations have demonstrated that acquisition times can be significantly shortened through the use of multiplexed LED patterns[46]. Traditionally, the reconstruction of objects in Fourier ptychography, hindered by the loss of phase information in intensity images, relies on iterative algorithms that demand significant computational resources. Recent advancements have illustrated that deep learning can serve as an effective substitute for these iterative processes, streamlining the reconstruction method[47]. In this single-shot imaging methodology, the conventional ptychography LED illumination pattern is optimized using deep learning techniques, allowing for the acquisition of fewer images without compromising the SBP (**Figure 3c**). Through the joint optimization of the LED illumination pattern and reconstruction network parameters, the deep learning model not only mitigates the impact of undersampling but also significantly reduces the acquisition time by a factor of e.g., 69[15].

Another illustration of undersampling data can be shown with the deep learning-assisted volumetric fluorescent microscopy system that uses a model called Recurrent-MZ[12]. This recurrent neural network (RNN)[48]-based volumetric image inference framework utilizes 2D images sparsely captured by a standard wide-field fluorescence microscope at arbitrary axial positions within the sample volume. Through a recurrent convolutional neural network, Recurrent-MZ incorporates 2D fluorescence information from a few axial planes within the sample to digitally reconstruct the sample volume over an extended DOF, as shown in **Figure 3d**. This approach significantly increases the imaging DOF of objective lenses and reduces the number of axial scans required to image the same sample volume, thereby advantageously undersampling data while maintaining imaging quality. These findings reveal that the Recurrent-MZ framework substantially enhances the DOF of a 63×/1.4NA objective lens, achieving a remarkable 30-fold decrease in the necessary axial scans for imaging the same sample volume. This RNN-based framework has also been applied to undersampling image data in holographic microscopy.[33]

Similarly, in the context of optical coherence tomography (OCT), the method known as Swept-Source OCT (SS-OCT)[14] leverages a deep learning-based image reconstruction approach to generate OCT images using undersampled spectral data. OCT is a non-invasive interferometric imaging technique capable of delivering 3D insights into the optical scattering characteristics of biological matter[49].This neural network-based SS-OCT approach eliminates spatial aliasing artifacts using less spectral data and without necessitating any hardware modifications to the optical setup. By training a deep neural network (DNN) on mouse embryo samples imaged by an SS-OCT system, researchers were able to blind-test the network's ability to reconstruct images using 2-3 fold undersampled spectral data, as shown in **Figure 3e**. The results showcase the network's potential to increase imaging speed without compromising image resolution or SNR.

The concept of compensating specifically for less *training data* with DNNs has also been applied to the process of hologram reconstruction[50]. One such technique uses a few-shot transfer learning style for holographic image reconstruction[13] that facilitates rapid generalization to new sample types using small datasets. Researchers pre-trained a convolutional recurrent neural network[33] on a dataset with three different types of samples and approximately 2000 unique sample FOVs, which served as the backbone model. By transferring only specific convolutional blocks of the pre-trained model, they dramatically reduced (by ~90%) the number of trainable parameters while achieving equivalent generalization to new samples. An example of this on lung tissue image reconstruction is shown in **Figure 3f**. Such an approach significantly accelerates convergence speed, reduces computation time, and improves generalization to new sample types, all while undersampling training data.

These methods highlight the strategic utilization of undersampling data in biophotonics and demonstrate how deep learning contributes to maximizing the advantages of this approach. Through deliberate compromise in data acquisition, these methodologies achieve enhanced imaging speed, reduced resource requirements, and minimized sample photodamage while maintaining or even improving imaging quality.

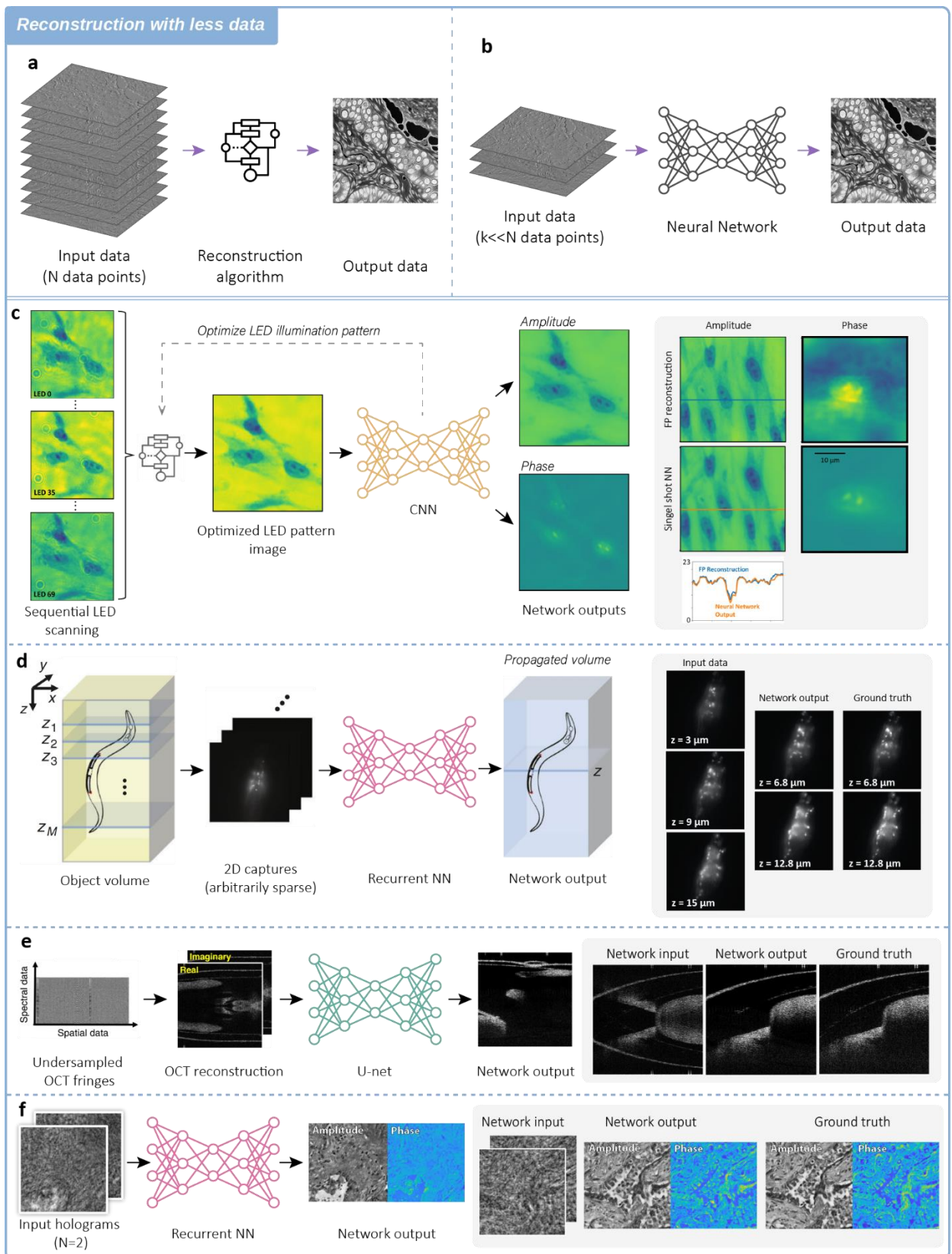

**Figure 3 Reconstruction with less data. a** Schematic representation of a typical reconstruction process. It consists of a dense set of input data and a standard algorithm. **b** With deep learning, a significant reduction from the traditional multi-layered stack is achieved for image reconstruction. This input is then fed into a neural network, which interprets and reconstructs the data. **c** Optimization of LED configuration using deep learning for Fourier ptychography with the resulting amplitude and phase components. An example from the evaluation dataset is provided for comparison, showcasing the phase component of the iterative FP reconstruction, which serves as the ground truth, alongside the output of the neural network, together with a cross-sectional analysis[15]. **d** The Recurrent-MZ volumetric imaging framework is illustrated through examples of 3D imaging of C. elegans, showcasing the initial input scans, the output processed by the network, and the established ground truths for comparison[12]. **e** The SS-OCT system acquires raw OCT fringes, from which the target image of the network is derived by directly reconstructing the original OCT fringes. By processing undersampled image through a trained network model, an OCT image free of aliasing is produced, closely aligning with the ground-truth. The provided example involves a 2× undersampled OCT image[14]. **f** Following a swift process of transfer learning, the RNN few shot hologram model demonstrates excellent generalization capabilities on test slides of new types of samples (lung tissue sections).

## 4 Improving image quality and throughput

This section presents various deep learning-enabled approaches to enhance the *quality* of the biophotonic data and the *throughput* of the overall system using modest, cost-effective or comprised equipment empowered by deep neural networks. In a similar vein to the previous methods discussed above, this approach leverages the power of neural networks to transform relatively suboptimal imaging data into high-quality representations, crucial for accurate biological analysis, all while forgoing some aspects of the optical hardware, including power, cost and form-factor.

As depicted in **Figure 4a**, traditional imaging systems using simplified devices invariably produce images of comparatively low quality in terms of SNR, spatial resolution, aberrations and DOF. Rather than having to rely on hardware-intensive setups to achieve first-rate results, it is possible to compensate for these deficiencies using DNNs. **Figure 4a** showcases the application of a neural network to process images captured from a cost-effective optical microscope. Here, the network acts on a single low-quality image, eliminating the need for multiple captures and complex optical systems, and outputs an image that closely resembles one obtained from a high-end benchtop microscope.

With respect to image quality, maintaining a high SNR[51] is of paramount importance. Achieving a higher SNR is critical for improving the sensitivity and accuracy of imaging techniques, especially in challenging conditions or when dealing with low light levels. Deep learning has emerged as a potent tool to augment SNR in such circumstances, leading to more reliable and informative imaging outcomes. A notable example of such an approach has been utilized on a type of mobile-phone microscopy[16]. While mobile phones have enabled cost-effective imaging technologies, their optical interfaces may introduce distortions/aberrations in imaging microscopic specimens, tampering with the SNR and image quality. Deep learning networks can correct these spatial and spectral aberrations, producing high-resolution, denoised, and color-corrected images that match the performance of benchtop microscopes with high-end, diffraction limited objective lenses. This method standardizes optical images for clinical and biomedical applications, augmenting SNR and overall image quality (**Figure 4b**).

A similar strategy was also presented in the paper titled "Deep learning enables fast, gentle STED microscopy" [21]. In the realm of STED microscopy[52], a technique that resolves features beyond the diffraction limit, the realization of super-resolution often comes at the cost of increased photobleaching and photodamage due to the necessity of high-intensity illumination. The use of deep learning in this paper aligns with the strategy of intentionally reducing pixel dwell time— thus sacrificing SNR and potentially image clarity—to improve the speed of imaging and reduce damage to biological samples, all while compensating for the sacrificed metrics using deep learning, as shown in **Figure 4c**.

Another technique that is coupled with a similar power saving process is stimulated Raman scattering microscopy (SRS) [53], a label-free imaging modality that offers chemical contrast based on the vibrational properties of molecules within a sample. It operates on the principle of Raman scattering, where incident light interacts with the molecular vibrations of the sample, leading to a shift in the energy of the scattered light. Deep learning has now been integrated with this technique to deliver a promising solution to significantly improve the SNR of SRS images [19]. As depicted in **Figure 4d,** a U-Net is trained to denoise SRS images of coronal mouse brain sections acquired with low SNR. The trained denoiser model also demonstrates external generalization to different imaging conditions, such as varying zoom and imaging depth, and augmenting SNR across various scenarios.

Another method involving overlapping FOVs [17] effectively broadens the throughput of a microscopic imaging system by addressing the inherent limitation of the SBP in conventional microscopes. In traditional settings, the SBP requirements of a microscope hinders the capability to process wide areas quickly and efficiently without sacrificing spatial details. This overlapped imaging system [17] includes a multi-lens array that circumvents the SBP bottleneck by capturing stacked images containing more information in a single snapshot, which can then be intricately processed and analyzed by an optimized machine learning model. This increases the throughput of the imaging process by a factor proportional to the number of FOVs that are integrated, allowing for a more efficient analysis of specimens, which is critical in biomedical research and disease diagnosis. This approach starts by lighting up different independent sample FOVs with LEDs. These are then imaged simultaneously through a multi-lens array onto a collective sensor, creating an overlapped composite image. To analyze this aggregate image, a CNN is designed to pinpoint and recognize distinct features or objects within it. **Figure 4e** shows an instance of this method, demonstrated through the model's ability to locate a target blood cell from an image of 2 overlapped FOVs. This technique directly aligns with the strategy of using deep learning to compensate for and correct the compromised elements of biophotonic imaging setups, facilitating advancements in imaging capabilities. This approach not only augments the throughput of microscopic analysis but also exemplifies the major impact of deep learning in expanding the operational envelope of conventional biophotonic imaging methods. If the composite image were to be unraveled into its individual FOV constituents using DNNs, this capability could be used to significantly enhance various detection processes in different sample types, such as tissue sections.

Lastly, an imaging configuration making use of a low light source in structured illumination microscopy (SIM) [18] showcases how deep learning improves SNR when imaging under extremely dim conditions. SIM, which works by illuminating the sample with patterned light, typically in the form of stripes or grids, and capturing multiple images as the pattern is shifted and rotated, thereby surpassing the optical diffraction limit, typically requiring intense illumination and multiple acquisitions [54]. Deep learning facilitates the production of high-resolution, denoised images of faintly illuminated samples, as shown in **Figure 4f** with microtubules. By enabling imaging with at least 100× fewer photons and 5x fewer raw data acquisitions (using fewer phase patterns in the illumination), this technique significantly boosts SNR, allowing for multi-color, live-cell super-resolution imaging with the added benefit of reduced photobleaching.

As highlighted through these examples, deep learning has significantly contributed to augmenting the SNR in biophotonics, enhancing the quality and reliability of imaging outcomes. These examples showcase how deep learning methods have effectively reduced noise, corrected distortions, and improved imaging volume, ultimately enhancing the SNR or the overall throughput of biophotonic imaging.

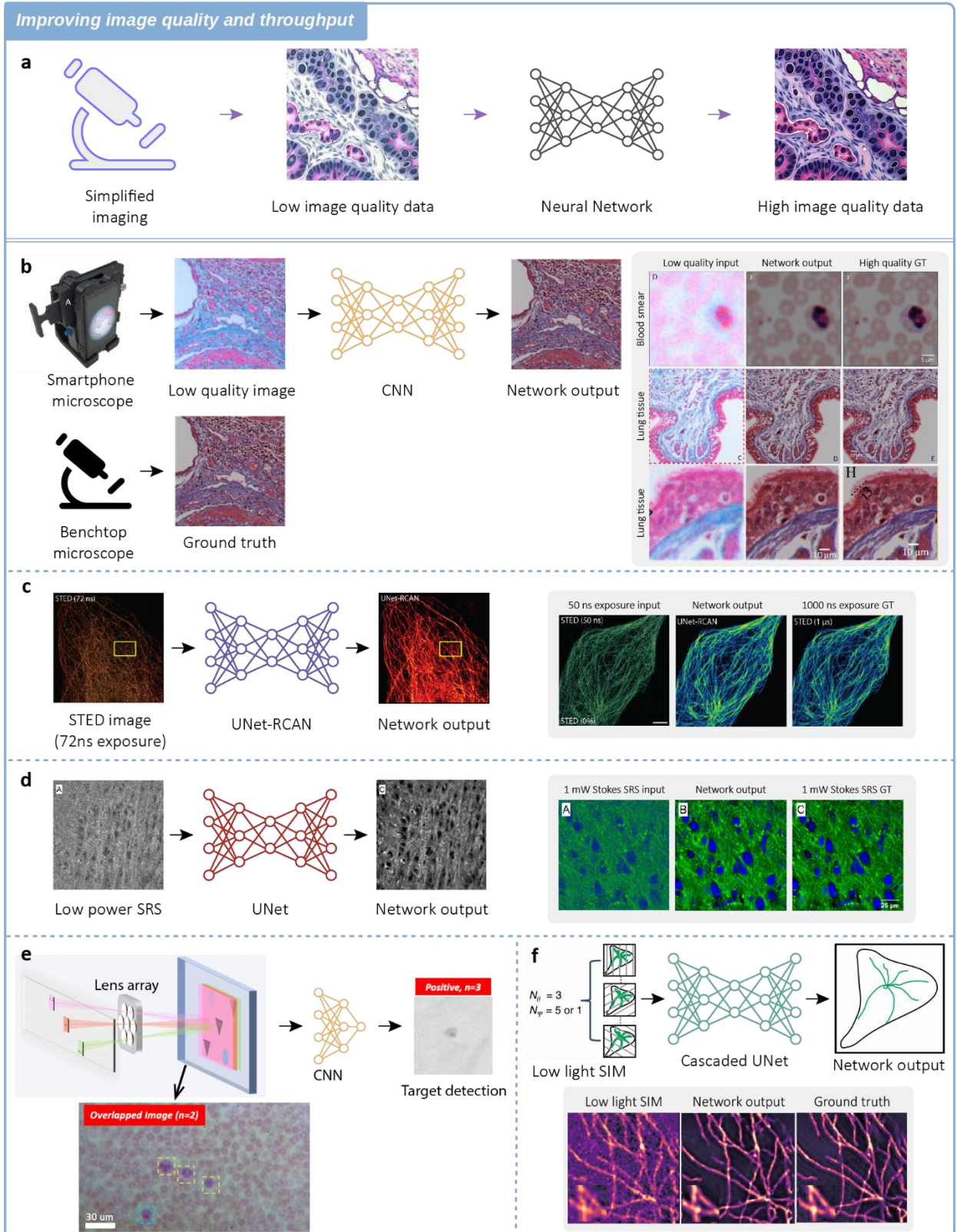

**Figure 4. Improving image quality and throughput**. **a** Schematic representation of a neural network-enabled pipeline for image quality improvement of data taken with simplified and/or inexpensive optics. **b** Deep learning enhanced mobile-phone microscopy

with a CNN trained to denoise, color-correct, and extend the depth of field with examples of blood smears and lung tissue sections. **c** Low exposure STED SNR enhancement through UNet-RCAN. The example shown compares noisy images (exposure time of 50 ns), ground-truth images (exposure time of 1 μs), and images processed by UNet-RCAN on β-tubulin (STAR635P) in U2OS cells[21]. **d** Reconstruction of low power (LP) SRS coronal mouse brain images and deep learning denoised versions, as well as two-color (lipids-green, proteins-blue) SRS images of a coronal mouse-brain slice with the ground truth as high power (HP) SRS images[19]. **e** The process of overlapped microscopy imaging involves illuminating various independent FOVs of samples using LEDs, followed by capturing these through a multi-lens array onto a shared sensor, resulting in an overlapped composite image. A CNN-based analysis framework is applied to detect and identify specific features within this composite image. This technique is exemplified by the model finding a target from an overlap of 3 images[17]. **f** Example of the low light SIM pipeline. For training the U-Net model, either fifteen (using three different illumination angles ($N_\theta$ =3) and five phase patterns ($N_\psi$ =5)) with faint illumination or three SIM raw data images (a single phase pattern for fewer raw data acquisitions) are employed as input, while high SNR SIM reconstructions serve as the ground truth. This approach is shown with examples on microtubules[18].

# 5 Discussion and future perspectives

The innovative integration of deep learning with biophotonic imaging represents a paradigm shift in bioimaging, offering a novel pathway to surpass traditional limitations and unlock new capabilities. This manuscript has illustrated how strategic compromises in measurement metrics, such as to the PSF, SNR, sampling density, and pixel resolution, can be effectively counterbalanced by designing and deploying specialized deep learning models. This approach not only recovers lost information, but also enhances imaging parameters critical for advanced biophotonic applications, such as the resolution, FOV, DOF, and SBP. The successful applications of these strategies across various biophotonic methods underscore the transformative potential of deep learning in bioimaging, pushing the boundaries of what is achievable in terms of temporal resolution, imaging speed, accessibility, and cost-effectiveness.

An intriguing prospect with regards to this compromise–compensate scheme is the potential to combine different imaging defects strategically to further enhance or expedite imaging processes. For example, researchers can leverage a compromised PSF alongside low SNR to accelerate image acquisition. By deliberately introducing these imperfections, it is possible to optimize imaging speed without significant loss of critical information. The creative fusion of defects could offer exciting prospects for real-time imaging in applications where rapid results are imperative.

The synthesis of different deep learning methods also opens a realm of interesting possibilities. Consider integrating overlapped microscopy with a GANscan acquisition strategy. Overlapping multiple FOVs on a single image sensor can significantly increase detection throughput. When fused with GANscan's already accelerated imaging capabilities, the result could revolutionize high-throughput imaging systems. There is, however, a legitimate concern that the adoption of these techniques in conjunction could face considerable pushback, for instance in scenarios where the precision and reliability of bioimaging are non-negotiable. The decision to leverage compromised imaging metrics for the sake of enhancing certain aspects of the imaging process, such as speed or FOV, necessitates a thorough and rigorous understanding of the trade-offs involved. It is essential to emphasize that the utility of these deep learning-enabled compromises is highly contingent upon the specific needs and constraints of the imaging task at hand.

In clinical diagnostics where the accuracy and reliability of imaging data are critical for patient care, the acceptance of assorted optical flaws to expedite imaging processes might not always be appropriate. Conversely, in research settings where speed and scalability of imaging are more crucial, such compromises might be more readily welcomed. This underscores the importance of system-specific considerations in the application and wide adoption of these advanced imaging techniques. Therefore, it is imperative for researchers and practitioners to critically evaluate the potential benefits and limitations (including potential hallucinations) of deep learning-enhanced bioimaging methods within their specific contexts. Understanding the precise requirements of their applications will allow the users to make more informed decisions about when and how to incorporate these innovative techniques, ensuring that the advancements in biophotonic imaging truly meet the nuanced demands of their work without unnecessarily sacrificing the quality or fidelity of the imaging data.

Finally, the path to Food and Drug Administration (FDA) approval with any of these systems remains a challenging milestone if the proposed systems are aimed to be used for diagnosing patients. Regulatory authorities often scrutinize compromised data, as it may raise concerns about unreliable medical results. However, the exponential growth in deep learning functionality and the vast wealth of data that is more and more at our disposal have the potential to address these concerns robustly. Deep learning algorithms, when rigorously validated and transparently documented, may prove, in time, their full reliability and safety. Furthermore, the generalization of AI models to learn from diverse datasets and adapt to different imaging conditions could mitigate the risks associated with compromised data. With the right protocols, standards, and clinical trials in place, we just may be able to navigate the regulatory landscape and demonstrate the immense benefits of deep learning in biophotonics without compromising on medical integrity.

As we stand on the cusp of this rapidly advancing field, it is clear that the future holds immense promise for myriad further innovations and breakthroughs.

## Disclosures

The authors declare no conflicts of interest.

## Author Contributions

A.O., M.J.F., and P.C.C. conceived the idea. P.C.C, M.J.F., Ç.I., and L.H. prepared the figures. All authors wrote the manuscript.

## Acknowledgments

Ozcan Lab acknowledges the support of NIH P41, The National Center for Interventional Biophotonic Technologies and the NSF Biophotonics program.

## References

1      Prasad, P. N. *Introduction to biophotonics*.  (John Wiley & Sons, 2004).

2      Marcu, L., Boppart, S. A., Hutchinson, M. R., Popp, J. & Wilson, B. C. Biophotonics: the big picture. *Journal of biomedical optics* **23**, 021103-021103 (2018).

3      Tian, L. *et al.* Deep learning in biomedical optics. *Lasers in Surgery and Medicine* **53**, 748-775 (2021).

4      Pradhan, P., Guo, S., Ryabchykov, O., Popp, J. & Bocklitz, T. W. Deep learning a boon for biophotonics? *Journal of Biophotonics* **13**, e201960186 (2020).

5    Icha, J., Weber, M., Waters, J. C. & Norden, C. Phototoxicity in live fluorescence microscopy, and how to avoid it. *BioEssays* **39**, 1700003 (2017).

6    Diaspro, A., Chirico, G., Usai, C., Ramoino, P. & Dobrucki, J. Photobleaching. *Handbook of biological confocal microscopy*, 690-702 (2006).

7    Demchenko, A. P. Photobleaching of organic fluorophores: quantitative characterization, mechanisms, protection. *Methods and applications in fluorescence* **8**, 022001 (2020).

8    Luo, Y., Huang, L., Rivenson, Y. & Ozcan, A. Single-shot autofocusing of microscopy images using deep learning. *ACS Photonics* **8**, 625-638 (2021).

9    Yang, X. *et al.* Deep-learning-based virtual refocusing of images using an engineered point-spread function. *ACS Photonics* **8**, 2174-2182 (2021).

10   Fanous, M. J. & Popescu, G. GANscan: continuous scanning microscopy using deep learning deblurring. *Light: Science & Applications* **11**, 265 (2022). https://doi.org:10.1038/s41377-022-00952-z

11   Chen, H., Huang, L., Liu, T. & Ozcan, A. eFIN: Enhanced Fourier Imager Network for generalizable autofocusing and pixel super-resolution in holographic imaging. *arXiv preprint arXiv:2301.03162* (2023).

12   Huang, L., Chen, H., Luo, Y., Rivenson, Y. & Ozcan, A. Recurrent neural network-based volumetric fluorescence microscopy. *Light: Science & Applications* **10**, 62 (2021).

13   Huang, L., Yang, X., Liu, T. & Ozcan, A. Few-shot transfer learning for holographic image reconstruction using a recurrent neural network. *APL Photonics* **7**, 070801 (2022).

14   Zhang, Y. *et al.* Neural network-based image reconstruction in swept-source optical coherence tomography using undersampled spectral data. *Light: Science & Applications* **10**, 155 (2021).

15   Cheng, Y. F. *et al.* Illumination pattern design with deep learning for single-shot Fourier ptychographic microscopy. *Optics express* **27**, 644-656 (2019).

16   Rivenson, Y. *et al.* Deep learning enhanced mobile-phone microscopy. *Acs Photonics* **5**, 2354-2364 (2018).

17   Yao, X. *et al.* Increasing a microscope's effective field of view via overlapped imaging and machine learning. *Optics Express* **30**, 1745-1761 (2022). https://doi.org:10.1364/OE.445001

18   Jin, L. *et al.* Deep learning enables structured illumination microscopy with low light levels and enhanced speed. *Nature Communications* **11**, 1934 (2020). https://doi.org:10.1038/s41467-020-15784-x

19   Manifold, B., Thomas, E., Francis, A. T., Hill, A. H. & Fu, D. Denoising of stimulated Raman scattering microscopy images via deep learning. *Biomedical optics express* **10**, 3860-3874 (2019).

20   Pinkard, H., Phillips, Z., Babakhani, A., Fletcher, D. A. & Waller, L. Deep learning for single-shot autofocus microscopy. *Optica* **6**, 794-797 (2019).

21   Ebrahimi, V. *et al.* Deep learning enables fast, gentle STED microscopy. *bioRxiv*, 2023.2001.2026.525571 (2023).

22   Goodfellow, I. *et al.* Generative adversarial networks. *Communications of the ACM* **63**, 139-144 (2020).

23   Shi, X. *et al.* Convolutional LSTM network: A machine learning approach for precipitation nowcasting. *Advances in neural information processing systems* **28** (2015).

24   Zhang, Y. *et al.* in *Proceedings of the European conference on computer vision (ECCV).* 286-301.

25   Rossmann, K. Point spread-function, line spread-function, and modulation transfer function: tools for the study of imaging systems. *Radiology* **93**, 257-272 (1969).

26   Jouchet, P., Roy, A. R. & Moerner, W. Combining deep learning approaches and point spread function engineering for simultaneous 3D position and 3D orientation measurements of fluorescent single molecules. *Optics Communications* **542**, 129589 (2023).

27    Astratov, V. N. *et al.* Roadmap on Label‐Free Super‐Resolution Imaging. *Laser & Photonics Reviews* **17**, 2200029 (2023).

28    Nehme, E. *et al.* DeepSTORM3D: dense 3D localization microscopy and PSF design by deep learning. *Nature methods* **17**, 734-740 (2020).

29    Vaquero, D., Gelfand, N., Tico, M., Pulli, K. & Turk, M. in *2011 IEEE Workshop on applications of computer vision (WACV).* 511-518 (IEEE).

30    Pavani, S. R. P. *et al.* Three-dimensional, single-molecule fluorescence imaging beyond the diffraction limit by using a double-helix point spread function. *Proceedings of the National Academy of Sciences* **106**, 2995-2999 (2009).

31    Wang, H. *et al.* Deep learning enables cross-modality super-resolution in fluorescence microscopy. *Nature methods* **16**, 103-110 (2019).

32    Wu, Y. *et al.* Extended depth-of-field in holographic imaging using deep-learning-based autofocusing and phase recovery. *Optica* **5**, 704-710 (2018).

33    Huang, L. *et al.* Holographic image reconstruction with phase recovery and autofocusing using recurrent neural networks. *ACS Photonics* **8**, 1763-1774 (2021).

34    Huang, L., Chen, H., Liu, T. & Ozcan, A. Self-supervised learning of hologram reconstruction using physics consistency. *Nature Machine Intelligence* **5**, 895-907 (2023).

35    Pirone, D. *et al.* Speeding up reconstruction of 3D tomograms in holographic flow cytometry via deep learning. *Lab on a Chip* **22**, 793-804 (2022).

36    Park, J. *et al.* Revealing 3D cancer tissue structures using holotomography and virtual hematoxylin and eosin staining via deep learning. *bioRxiv*, 2023.2012. 2004.569853 (2023).

37    Barbastathis, G., Ozcan, A. & Situ, G. On the use of deep learning for computational imaging. *Optica* **6**, 921-943 (2019).

38    Situ, G. Deep holography. *Light: Advanced Manufacturing* **3**, 278-300 (2022).

39    Kakkava, E. *et al.* Imaging through multimode fibers using deep learning: The effects of intensity versus holographic recording of the speckle pattern. *Optical Fiber Technology* **52**, 101985 (2019).

40    Park, J. *et al.* Artificial intelligence-enabled quantitative phase imaging methods for life sciences. *Nature Methods* **20**, 1645-1660 (2023).

41    Chen, H., Huang, L., Liu, T. & Ozcan, A. Fourier Imager Network (FIN): A deep neural network for hologram reconstruction with superior external generalization. *Light: Science & Applications* **11**, 254 (2022).

42    Lepage, G., Bogaerts, J. & Meynants, G. Time-delay-integration architectures in CMOS image sensors. *IEEE Transactions on Electron Devices* **56**, 2524-2533 (2009).

43    Ren, Z., Xu, Z. & Lam, E. Y. Learning-based nonparametric autofocusing for digital holography. *Optica* **5**, 337-344 (2018).

44    Konda, P. C. *et al.* Fourier ptychography: current applications and future promises. *Optics express* **28**, 9603-9630 (2020).

45    Zheng, G., Horstmeyer, R. & Yang, C. Wide-field, high-resolution Fourier ptychographic microscopy. *Nature photonics* **7**, 739-745 (2013).

46    Tian, L., Li, X., Ramchandran, K. & Waller, L. Multiplexed coded illumination for Fourier Ptychography with an LED array microscope. *Biomedical optics express* **5**, 2376-2389 (2014).

47    Nguyen, T., Xue, Y., Li, Y., Tian, L. & Nehmetallah, G. Deep learning approach for Fourier ptychography microscopy. *Optics express* **26**, 26470-26484 (2018).

48    Grossberg, S. Recurrent neural networks. *Scholarpedia* **8**, 1888 (2013).

49    Podoleanu, A. G. Optical coherence tomography. *The British journal of radiology* **78**, 976-988 (2005).

50    Kim, M. K. Principles and techniques of digital holographic microscopy. *SPIE reviews* **1**, 018005 (2010).

51      Stelzer. Contrast, resolution, pixelation, dynamic range and signal‐to‐noise ratio: fundamental limits to resolution in fluorescence light microscopy. *Journal of Microscopy* **189**, 15-24 (1998).

52      Rittweger, E., Han, K. Y., Irvine, S. E., Eggeling, C. & Hell, S. W. STED microscopy reveals crystal colour centres with nanometric resolution. *Nature Photonics* **3**, 144-147 (2009).

53      Tipping, W., Lee, M., Serrels, A., Brunton, V. & Hulme, A. Stimulated Raman scattering microscopy: an emerging tool for drug discovery. *Chemical Society Reviews* **45**, 2075-2089 (2016).

54      Saxena, M., Eluru, G. & Gorthi, S. S. Structured illumination microscopy. *Advances in Optics and Photonics* **7**, 241-275 (2015).